\documentclass[10pt,letterpaper]{article}
\usepackage{opex3}

\begin{document}

%%%%%%%%%%%%%%%%%% title page information %%%%%%%%%%%%%%%%%%
\title{Real-time determination of laser beam quality by modal decomposition}

\author{Oliver A. Schmidt,$^{1,2}$ Christian Schulze,$^1$ Daniel Flamm,$^1$ \\
 Robert Br\"uning,$^1$ Thomas Kaiser,$^3$ Siegmund Schr\"oter,$^4$ \\
and Michael Duparr\'e$^{1,*}$ }

\address{$^1$Institute of Applied Optics, Friedrich-Schiller-University Jena, D-07743 Jena, Germany\\
$^2$now with the Max Planck Institute for the Science of Light, D-91058 Erlangen, Germany\\
$^3$Institute of Applied Physics, Friedrich-Schiller-University Jena, D-07743 Jena, Germany\\
$^4$Institute of Photonic Technology, Albert-Einstein-Strasse 9, D-07745 Jena, Germany}

\email{$^*$michael.duparre@uni-jena.de}  %% email address is required

%%%%%%%%%%%%%%%%%%% abstract and OCIS codes %%%%%%%%%%%%%%%%
%% [use \begin{abstract*}...\end{abstract*} if exempt from copyright]

\begin{abstract}
We present a real-time method to determine the beam propagation ratio $M^2$ of laser beams. The all-optical measurement of  modal amplitudes yields $M^2$ parameters conform to the ISO standard method. The experimental technique is simple and fast, which allows to investigate laser beams under conditions inaccessible to other methods.
\end{abstract} 
 % REPLACE WITH CORRECT OCIS CODES FOR YOUR ARTICLE

\textbf{This paper was published in Optics Express and is made available as an electronic reprint with the permission of OSA. The paper can be found at the following URL on the OSA website: 
\begin{center}
\texttt{http://www.opticsinfobase.org/oe/abstract.cfm?uri=oe-19-7-6741}.
\end{center}
Systematic or multiple reproduction or distribution to multiple locations via electronic or other means is prohibited and is subject to penalties under law.}\\

%%%%%%%%%%%%%%%%%%%%%%% References %%%%%%%%%%%%%%%%%%%%%%%%%

%\bibliography{literature}   %>>>> bibliography data in report.bib
%\bibliographystyle{osajnl}

%%%%%%%%%%%%%%%%%%%%%%%%%%  body  %%%%%%%%%%%%%%%%%%%%%%%%%%
\section{Introduction}
The characterization of laser beams is traditionally a basic task in optical science and performed for decades now. However, the existence of different archaic approaches like the variable aperture or moving knife-edge method, which are known to produce deviant results, has shown the need for a standardization of the definition. Especially the question how  to reproducibly and reliably measure beam quality, which is very important from an user-oriented point of view, stimulated a lot of discussions~\cite{Siegman1998}. The ISO standard 11146-1/2/3 has brought a great unification by defining all relevant quantities of laser beams including instructions how to perform the measurements conform to the ISO standard~\cite{ISO11146}.
In the general case of so-called general astigmatic beams this approach relies on the determination of ten independent parameters, which are the second order moments of the Wigner distribution function.
From these, three quantities can easily be derived according to ISO 11146-2: the beam propagation ratio simply called $M^2$ parameter, the intrinsic astigmatism, and the twist parameter~\cite{ISO11146}. Especially the $M^2$ factor has become a well accepted measure for beam quality in the laser community by now---although it has to be used with care.

Siegman pointed out that the use of measurements not conform to the ISO standard will result in values for $M^2$, which are not comparable to each other~\cite{Siegman1998}. This is an important fact since other techniques as the aforementioned knife-edge method are still in use for one reason: their simplicity. As stated above, a measurement of a general astigmatic beam, which is fully conform to the ISO standard, requires the measurement of ten second order moments of the Wigner distribution function and, thus, is experimentally cumbersome and slow.

Fortunately, due to their high symmetry most laser beams of practical interest require less than these ten independent parameters to be measured for a complete characterization: Most lasers emit beams that are simple astigmatic or even stigmatic because of their resonator design. In this case the $M^2$ determination is based on a caustic measurement that includes the determination of beam width as a function of propagation distance. To do this, the beam width has to be determined at least at 10 positions along the propagation axis, completed by a hyperbolic fit. According to ISO 11146-1, the beam width determination has to be carried out using the three spatial second order moments of the intensity distribution~\cite{ISO11146}.

Despite its experimental simplicity, the caustic measurement still is quite time-consuming and requires a careful treatment of background intensity, measuring area, and noise, which makes high demands on the temporal stability of a cw laser or the pulse-to-pulse stability of a pulsed source. Caustic measurements are therefore unsuitable to characterize the fast dynamics of a laser system. To react on the need for a faster and more detailed analysis, several other methods for laser beam characterization were presented such as Shack-Hartmann wavefront analysis~\cite{Schafer2002}, measurement of the complete 4D Wigner distribution function~\cite{Eppich2005}, or the use of diffraction gratings~\cite{Lambert2004}.

In this paper, we investigate the possibility to monitor the beam quality based on a decomposition of the laser beam into its constituent eigenmodes. Many laser resonators possess rectangular or circular symmetry. In this case, the laser beam emerging from the resonator can completely be described as a superposition of the well-known Hermite-Gaussian or Laguerre-Gaussian modes~\cite{Siegman1986}. From the point of view of beam quality, they represent a natural choice of description since the fundamental Gaussian mode has the lowest possible value of $M^2=1$. Any deviation from the ideal diffraction-limited Gaussian beam profile can be attributed to the contribution of higher order modes, leading to $M^2>1$. Here, any excited higher order mode contributes with a certain value to the beam propagation parameter. This value can be readily calculated for every mode in a way which is conform to the ISO standard~\cite{Saghafi1998}. Therefore, performing a modal decomposition of a given beam enables us to determine a value for $M^2$ which is compatible to the one resulting from a caustic measurement.

To determine the modal weights necessary for the decomposition, various methods have been suggested such as coherence measurements~\cite{Tervonen1989}, intensity recordings~[9--11], %~\cite{Cutolo1995,Santarsiero1999,Xue2000}, 
or the use of a ring resonator~\cite{Andermahr2008}. Again, one has to see the required experimental or numerical effort that makes those methods unsuitable for the intended task of real-time characterization. Hence, we suggest to use the correlation filter technique based on computer-generated holograms (CGHs)~\cite{Soifer1994,Duparre2005}. 
Recently we successfully applied this method to characterize the field produced by multimode optical fibers and to determine the polarization states of different modes~\cite{Kaiser2009a,Flamm2010a}.

Here we will show that the correlation filter method yields $M^2$ values conform to the ISO standard but with diverse advantages. The experimental realization is simple and yields information about the beam quality in real-time in contrast to other approaches. It relies on a decomposition of the electromagnetic field into the spatial modes of the resonator generating the laser beam. Since our investigated laser cavity possesses rectangular symmetry, Hermite-Gaussian modes are used in this work, but represent no restriction of the method. Any complete set of suitable modes may be used in other cases.

From the measured modal weights, we calculate the beam propagation factor $M^2$ and compare our results to values obtained from traditional caustic measurements as defined by the ISO standard. Moreover, we demonstrate the on-line monitoring of the modal spectrum as well as the $M^2$ factor while realigning the resonator in real-time.

\section{Modal decomposition}
\label{sec:fund}

\subsection{Expansion into Hermite-Gaussian modes}

Laser resonators with rectangular geometry generate superpositions of nearly Hermite-Gaussian (HG) modes  with the field distribution~\cite{Kogelnik1966}
\begin{equation}
\mathrm{HG}_{mn}(x,y) = \frac{1}{w_0} \sqrt{ \frac{2}{2^{m+n}\,\pi \, m! \, n!}} \ \mathrm{H}_m\! \left(\frac{ \sqrt 2 }{w_0} \,x\right) \mathrm{H}_n\! \left(\frac{ \sqrt 2 }{w_0} \,y\right) \exp\! \left(- \frac{x^2+y^2}{w_0^2}\right)\ ,
\end{equation}
where $w_0$ is the waist radius of the fundamental mode $\mathrm{HG}_{00}$ and $\mathrm{H}_l$ denotes the Hermite polynomial of order $l$.

Hermite-Gaussian modes are one set of orthogonal eigenfunctions of the scalar Helmholtz wave equation. Due to the completeness of this eigenfunction set, an arbitrary transverse wave field $U$ can be expanded into a superposition of HG modes~\cite{Siegman1986}
\begin{equation}
U(x,y)=\sum_{m=0}^{\infty}\sum_{n=0}^{\infty} c_{mn}  \mathrm{HG}_{mn}(x,y) \ , \quad c_{mn} = \int\!\!\!\!\!\int_{-\infty}^\infty \mathrm{HG}_{mn}^* \, U \,\mathrm{d} x  \,\mathrm{d} y \ ,
\label{eq:dec}
\end{equation}
where the asterisk denotes complex conjugation. The complex-valued expansion coefficients $c_{mn}= \rho_{mn}  \mathrm{e}^{\mathrm{i}\phi_{mn}} $ include modal amplitudes $\rho_{mn}=\left|c_{mn}\right|$ and phases $\phi_{mn}=\mathrm{arg}(c_{mn})$. Using normalized fields with unit power $P=\int\!\!\!\int_{-\infty}^\infty \left|U\right|^2 \,\mathrm{d} x \,\mathrm{d} y =1$, the squared amplitudes $\rho_{mn}^2$ represent the relative power of the $\mathrm{HG}_{mn}$ mode since $\sum_{m, n} \rho_{mn}^2=1$.

\subsection{Beam propagation ratio}

According to the ISO standard~\cite{ISO11146}, the beam propagation ratio for simple astigmatic beams is defined by 
\begin{equation}
M^2_{x,y}=\frac{\pi}{\lambda}\frac{d_{x,y} \theta_{x,y}}{4} \ ,
\label{eq:M2sa}
\end{equation}
where $d$ is the beam diameter and $\theta$ denotes the divergence angle. These parameters are based on second order moments $\sigma^2$. For instance, the beam diameter in $x$-direction reads as~\cite{ISO11146}
\begin{equation}
d_x = \sqrt{8} \left[ \sigma_{x}^2+\sigma_{y}^2 + \gamma \sqrt{\left(\sigma_{x}^2-\sigma_{y}^2 \right)^2+4\left(\sigma_{xy}^2 \right)^2} \right]^{1/2} , \quad 
\gamma=\frac{\sigma_{x}^2-\sigma_{y}^2}{\left|\sigma_{x}^2-\sigma_{y}^2\right|} \ .
\label{eq:width} 
\end{equation}

Using HG modes, the beam quality from Eq.~(\ref{eq:M2sa}) simplifies to
\begin{equation}
M^2_{x,y}=\frac{d_{x,y}^2}{4w_0^2} \ .
\label{eq:M2HG}
\end{equation}
In other words, the ratio of the beam waist diameter to the one of the fundamental Gaussian beam determines the $M^2$ factor. 

The laser used for experimental demonstration (manufactured by Smart Laser Systems) generates simple astigmatic beams composed of different HG$_{m0}$ modes with higher order modes in one direction~\cite{Laabs1996}. In general, different modes of a resonator possess slightly dissonant frequencies~\cite{Kogelnik1966}. Only modes with $p=m+n=const$ belonging to one \emph{mode group} have the same frequency and, thus, temporally stable intermodal phase differences. 
Due to the huge camera integration time compared to the beating period, the interference terms between modes of different mode groups vanish, whereas modes of one and the same mode group contribute coherently to a recorded intensity distribution. Furthermore, a rotation of the principal axis of one mode towards the laboratory system can be described by a superposition of HG modes of the associated mode group---similar to the expansion of Laguerre-Gaussian modes into HG modes~\cite{Siegman1986}. Consequently, the total intensity distribution is the sum of intensity contributions from single mode groups
\begin{equation}
I = \left|U\right|^2 = \sum_{p=0}^{\infty}I_p \quad\mathrm{with}\quad I_p=\sum_{k, m} \rho_{k,p-k}\rho_{m,p-m}  \, \mathrm{HG}_{k,p-k} \, \mathrm{HG}_{m,p-m} \ ,
\label{eq:intens}
\end{equation}
where $p$ denotes the $p^\mathrm{th}$ mode group and $I_p$ is the intensity formed by a superposition of the modes from mode group $p$.

Using the intensity profile from Eq.~(\ref{eq:intens}), the first order moments of the beam vanish and, thus, express that the beam centroid is on the optical axis. Hence, the spatial second order moments read as 
\begin{equation}
 \sigma_x^2=  \int\!\!\!\!\!\int_{-\infty}^\infty x^2 \, I(x,y) \,\mathrm{d} x \,\mathrm{d} y  = \frac{w_0^2}{4} \sum_{m, n} \, \rho_{mn}^2 \, (2m+1) \ ,
 \label{eq:sigmax2} 
\end{equation}
\begin{equation}
 \sigma_{xy}^2= \int\!\!\!\!\!\int_{-\infty}^\infty x y \, I(x,y) \,\mathrm{d} x \,\mathrm{d} y = \frac{w_0^2}{2} \sum_{m, n} \rho_{m+1,n} \, \rho_{m,n+1} \, \sqrt{m+1} \sqrt{n+1}  \ ,
 \label{eq:sigmaxy}
\end{equation}
and analogously for $\sigma_y^2$, where the identities~\cite{Szego1975}
\begin{equation}
 \int\!\!\!\!\!\int_{-\infty}^\infty x^2 \, \mathrm{HG}_{kl}  \mathrm{HG}_{mn} \,\mathrm{d} x \,\mathrm{d} y  
 =  \frac{w_0^2}{4} \delta_{ln}  \left[(2m+1) \delta_{km} +\sqrt{\mathrm{max}(k,m)}
 \sqrt{\mathrm{max}(k,m)-1} \, \delta_{\left|k-m\right|,2} \right] \ , 
\end{equation}
\begin{equation}
 \int\!\!\!\!\!\int_{-\infty}^\infty x y \, \mathrm{HG}_{kl}  \mathrm{HG}_{mn} \,\mathrm{d} x \,\mathrm{d} y  = 
 \frac{w_0^2}{4} \sqrt{\mathrm{max}(k,m)} \ \delta_{\left|k-m\right|,1} \sqrt{\mathrm{max}(l,n)} \ \delta_{\left|l-n\right|,1} \ , 
\end{equation}
and the Kronecker symbol $\delta_{ij}$ have been used.

Combining Eq.~(\ref{eq:width}) and (\ref{eq:M2HG}) with Eq.~(\ref{eq:sigmax2}) and (\ref{eq:sigmaxy}), the beam propagation ratio in $x$-direction reads as 
\begin{eqnarray}
\nonumber M_x^2 & = & \sum_{m, n} \, \rho_{mn}^2 \, (m+n+1)  \\ 
   & + & \gamma \left[ \left(\sum_{m, n} \, \rho_{mn}^2 \, (m-n) \right)^2+4\left(\sum_{m, n} \rho_{m+1,n} \, \rho_{m,n+1} \, \sqrt{m+1} \sqrt{n+1} \right)^2 \right]^{1/2}  \ .
\label{eq:M2}
\end{eqnarray}
To obtain $M_y^2$, the plus sign in front of $\gamma$ has to be replaced by a minus. Due to Eq.~(\ref{eq:M2}), the knowledge of the modal weights is sufficient to determine the beam propagation ratio $M^2$.

\section{Experiments}
\subsection{Measurement setup}

\begin{figure}[htb]
\centering \includegraphics[width=0.9\textwidth]{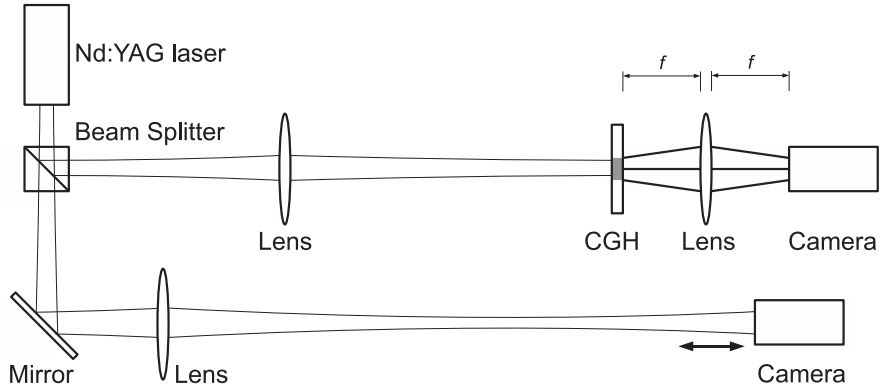}
\caption{Scheme of experimental setup. Upper branch: $M^2$ determination using a CGH. Lower branch: $M^2$ determination by a caustic measurement conform to the ISO standard.}
\label{fig:setup}
\end{figure}

The optical setup  in Fig.~\ref{fig:setup} consists of a laser source and two branches that enable two simultaneous but independent measurements of the $M^2$ parameter of the emerging beam. The Nd:YAG laser can produce different HG$_{mn}$ modes at $\lambda=1064\,$nm. The excited HG$_{mn}$ modes are restricted to $n=0$ since the pump light ($\lambda=808\,$nm) is coupled into the plane-concave resonator by a horizontally movable fiber~\cite{Laabs1996}. A beam splitter divides the beam and guides the replicas into the two branches for analysis.

In the first branch a lens images the laser beam waist onto an adapted CGH, which is an amplitude hologram consisting of $601\times601$ Lee cells with cell widths of 16\,$\mu$m. The far field intensity behind the CGH realized by a subsequent $2f$-setup is recorded by a CCD camera. This intensity pattern contains the information about the weights $\rho_{mn}^2$ of the HG modes since the complex conjugated modes are implemented in the CGH called \emph{correlation filter}~\cite{Kaiser2009a}. Hence, the correlation filter technique enables the determination of the $M^2$ factor in real-time using a computer-aided calculation of Eq.~(\ref{eq:M2}). 

The CGH for this experiment is designed to simultaneously analyze the amplitudes of 21 HG modes, i.e., all HG$_{mn}$ modes with $(m+n)\leq5$ are implemented taking into account a possible rotation of the resonator coordinate system. This correlation filter configuration corresponds to a truncation of all sums in the equations of section~\ref{sec:fund}.

In the second branch of the setup in Fig.~\ref{fig:setup}, an additional lens is used to analyze the caustic of the beam conform to the ISO standard~\cite{ISO11146}, which serves as a reference measurement.

\subsection{Measurement results}

\begin{figure}[htb]
\centering
\includegraphics[viewport=2.5cm 0cm 41cm 24cm,clip,height=0.4\textwidth]{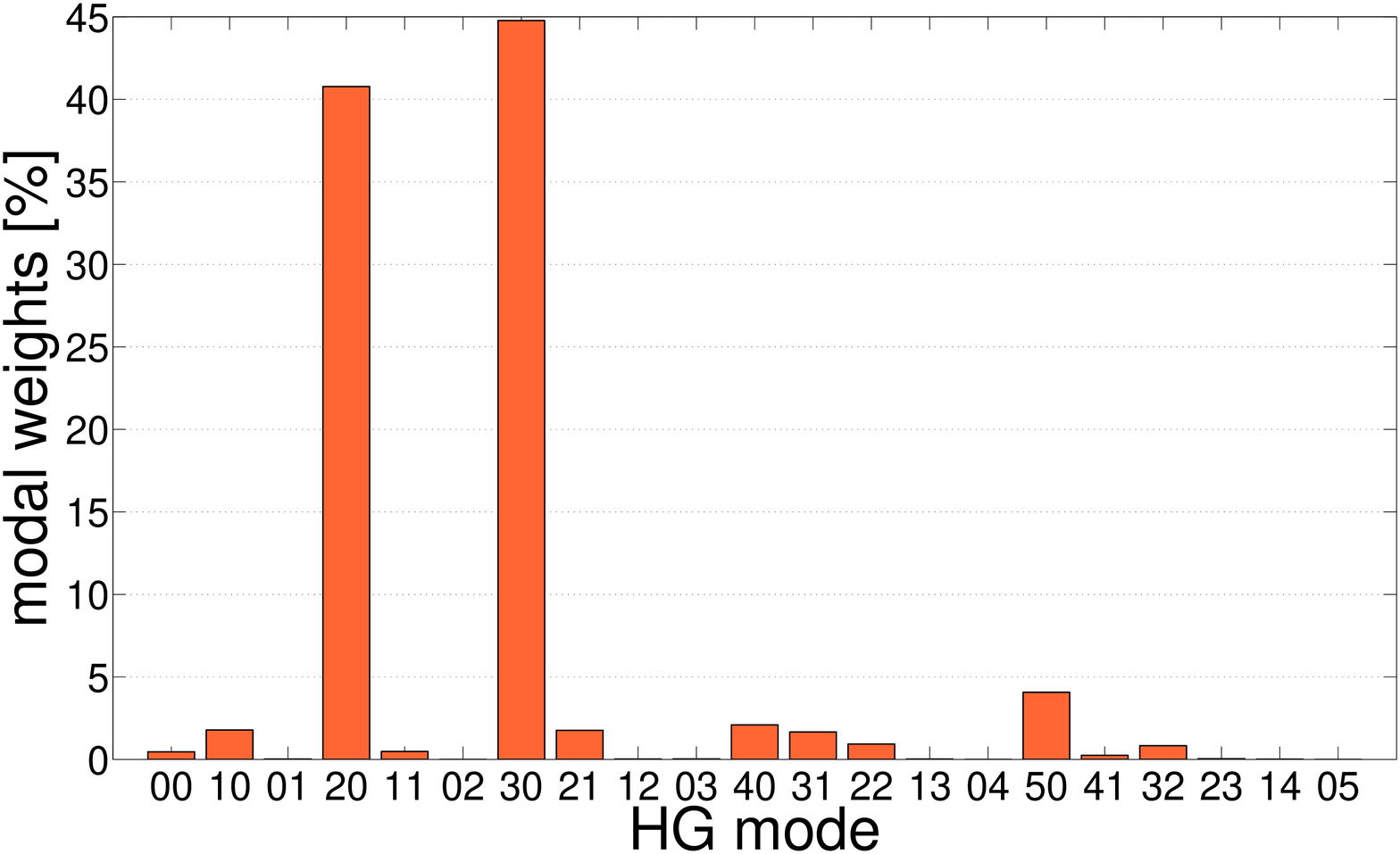} \qquad
\includegraphics[height=0.4\textwidth]{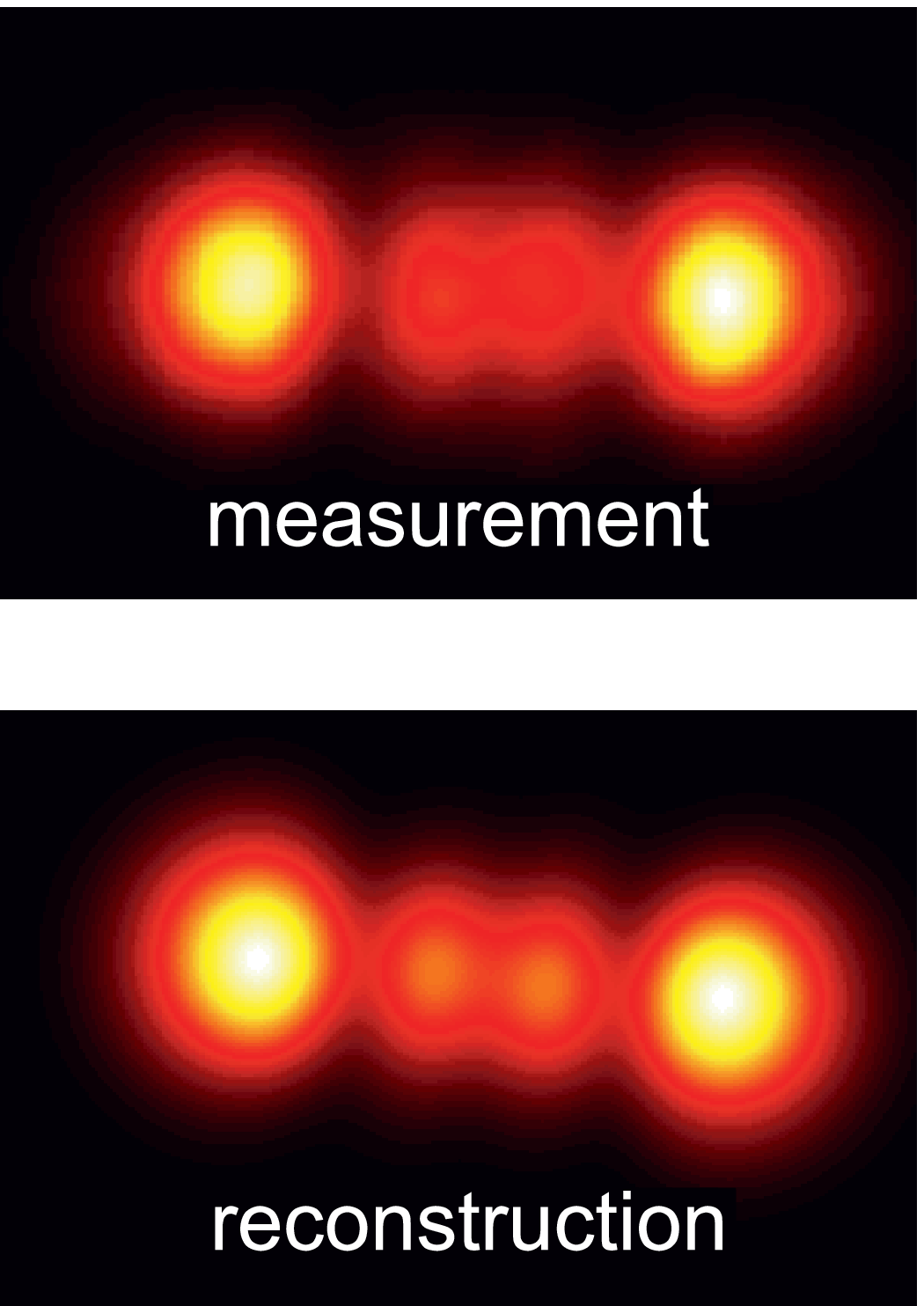} 
\caption{Modal spectrum of an arbitrarily chosen mode mixture (left) with corresponding measured and reconstructed near field intensity distributions (right).}
\label{fig:beams}
\end{figure}

To investigate the reliability of the CGH-based measurement procedure, a modal spectrum of an arbitrarily chosen beam consisting mainly of $\mathrm{HG}_{20}$ and $\mathrm{HG}_{30}$ is depicted in Fig.~\ref{fig:beams}. Using Eq.~(\ref{eq:intens}) the intensity distribution in the plane of the CGH can be reconstructed. Since the method will only be as good as it is able to reconstruct the investigated beam, the measured intensity distribution is compared to the reconstructed one by calculating their two-dimensional cross-correlation coefficient. A value of 1.0 denotes perfect match. Cross-correlation coefficients greater than 0.9 in all investigated cases indicate the excellent functionality of the method. In particular, the cross-correlation coefficient for the beam investigated in Fig.~\ref{fig:beams} is 0.98.
For the determination of $M^2$, the reconstruction of the intensity profile is not necessary as a matter of course and just shown here to illustrate the functionality of the method.

\begin{figure}[htb]
\centering 
\includegraphics[viewport=0cm 1cm 41cm 23cm,clip,width=0.49\textwidth]{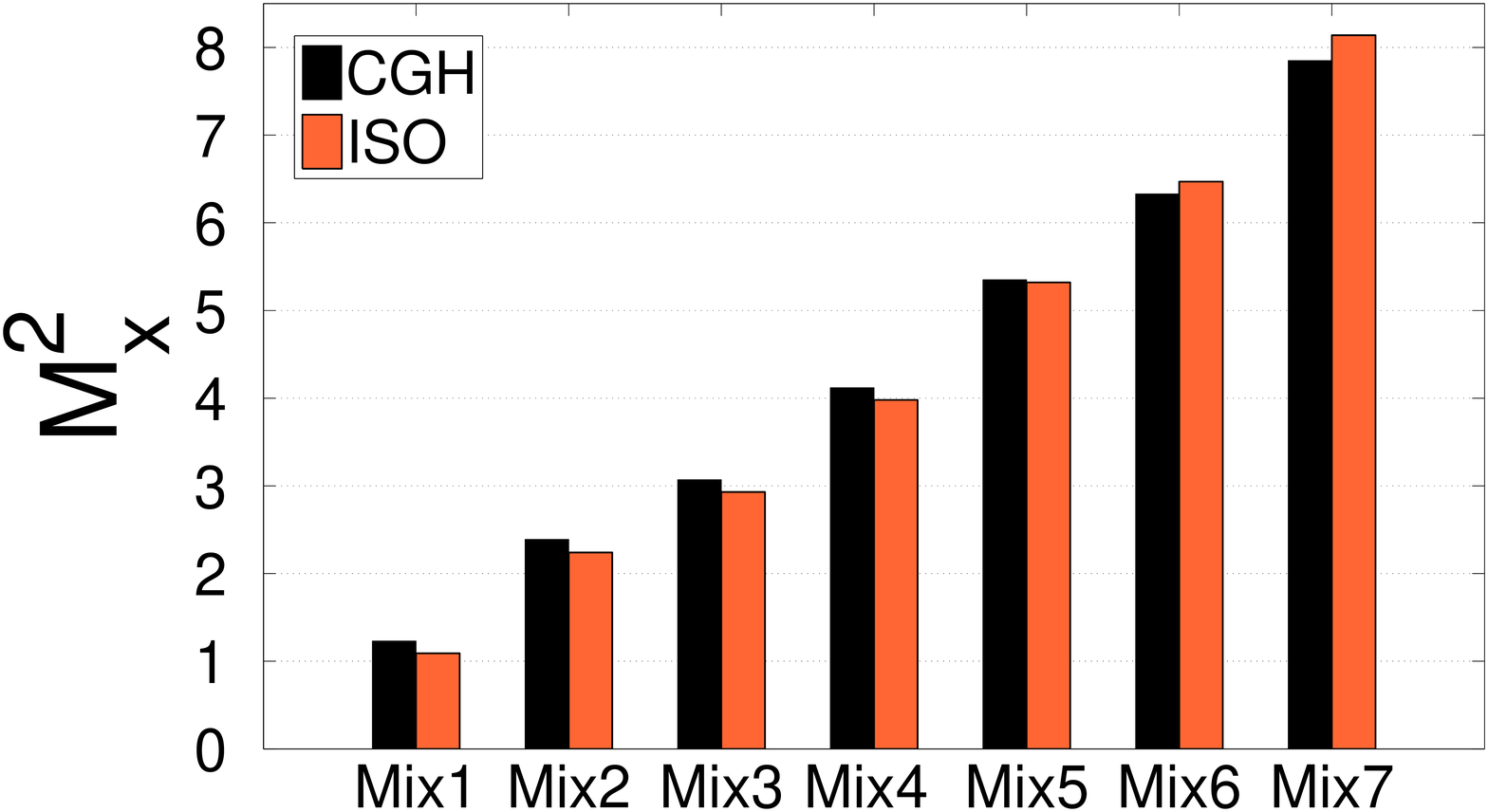} \ 
\includegraphics[viewport=0cm 1cm 41cm 23cm,clip,width=0.49\textwidth]{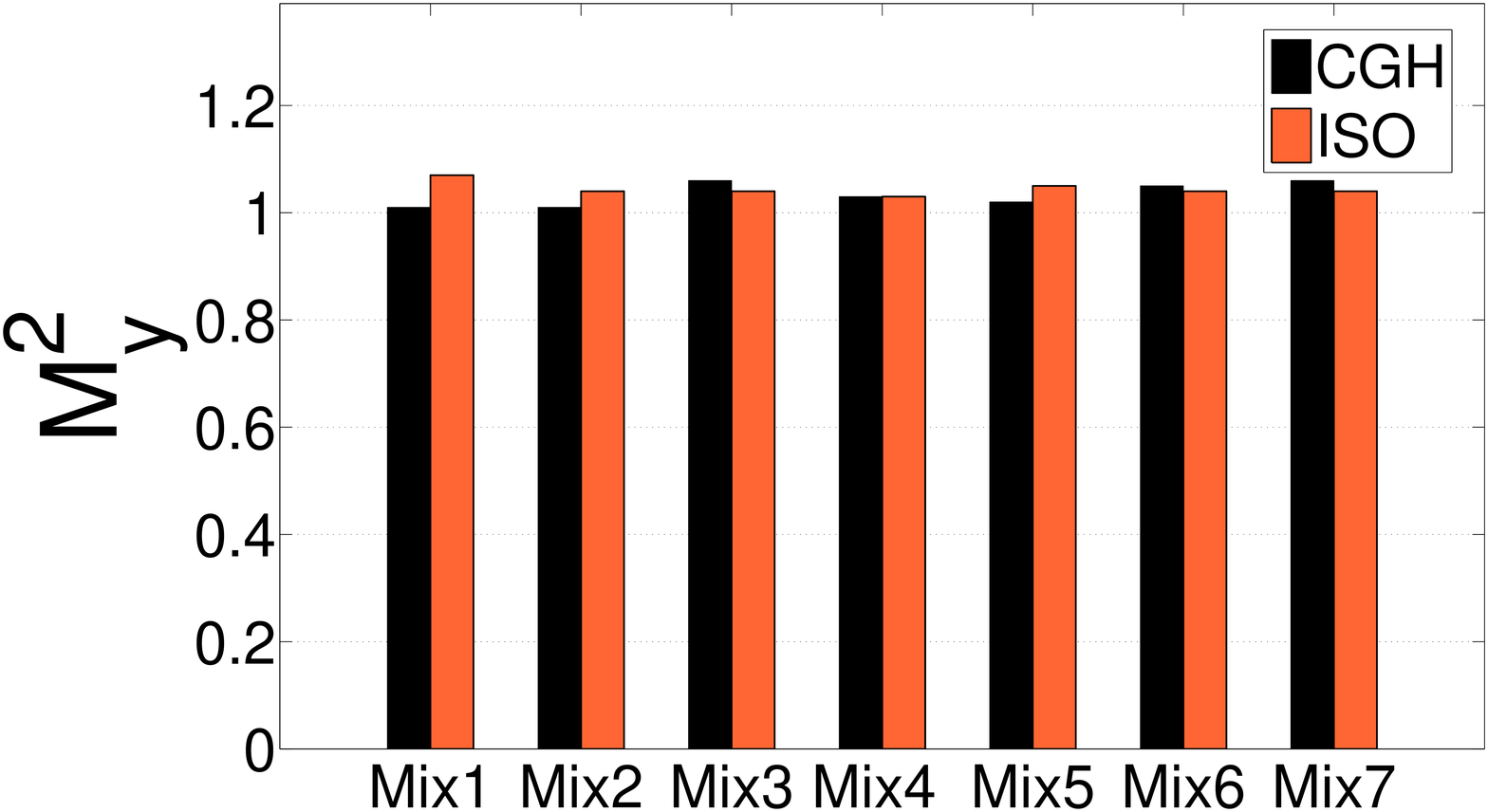}
\caption{Comparison of the $M^2_x$ and $M^2_y$ factors determined by the CGH technique and the ISO standard method, respectively.}
\label{fig:resultx}
\end{figure}

The results for the measurement of the beam propagation ratio using the correlation filter technique and the ISO standard method are compared in Fig. \ref{fig:resultx}. Seven different mode mixtures are selected and analyzed to test the accordance of the results of both techniques for beams with different $M^2$ factors. For instance, the beam in Fig.~\ref{fig:beams} is denoted as Mix\,6. As the bar charts in Fig.~\ref{fig:resultx} show, the results of both measurement techniques are in very good agreement. The $M_y^2$ parameters confirm the characteristic of the laser to emit only mixtures of $\mathrm{HG}_{m0}$ modes consistent with the theory. The slight deviations in the $M_y^2$ factors are caused by several reasons: First, the hologram needs to be placed accurately at the waist position and on the optical axis. Furthermore, the waist diameter has to match  the beam diameter the correlation filter is designed for. Any deviance from the ideal alignment as well as CCD noise affect the measured modal weights resulting in slight inaccuracies of the CGH-based beam quality determination. 

The deviation between the differently determined $M_y^2$ factors behaves statistically, whereas a tendency in the deviation of the $M_x^2$ values is recognizable: For mixtures with low $M_x^2$ factors such as Mix\,1 to Mix\,4, the $M_x^2$ values determined with the CGH are larger than the ones of the ISO conformable measurement. Here slight errors of the modal weights occur for higher order modes due to CCD noise producing an offset of the implemented higher order mode amplitudes, which enlarge the CGH-based $M_x^2$ factors. On the other hand, Mix\,6 and 7 mainly consist of modes from mode groups with $m+n$ near the truncation limit. Due to the finite number of implemented modes in the CGH, modal weights of HG$_{mn}$ modes with $(m+n)>5$, which would increase the beam quality $M_x^2$, are set to zero in the experiment. Hence, the missing amplitudes cause reduced $M_x^2$ factors. For Mix\,5 these two effects nearly compensate one another. 

In all analyzed cases, the maximum measured deviation is 13\% for $M_x^2$ and 5\% for $M_y^2$. This result demonstrates the functionality of the method, especially when beams with high beam quality are considered.

\begin{figure}[htb]
\center
\includegraphics[viewport=1.9cm 0.5cm 41cm 20cm,clip,width=0.9\textwidth]{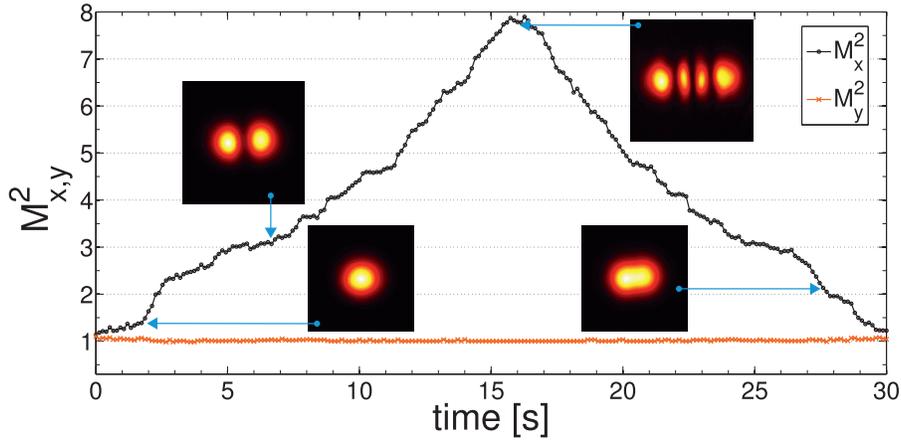}
\caption{$M_x^2$ and $M_y^2$ on a time scale of $30$\,s taken with a rate of $4$\,Hz. To vary the beam quality the laser resonator was tuned continuously. The insets depict the waist intensities at selected points in time.}
\label{fig:M2t}
\end{figure}

To demonstrate the real-time capability of the all-optical CGH technique, we continuously varied the alignment of the laser cavity and monitored the output. Figure~\ref{fig:M2t} illustrates an $M^2$ measurement on a time scale of 30\,s, where the cavity was initially aligned to produce the fundamental HG$_{00}$ mode. Then we changed the transverse position of the end-pumping fiber~\cite{Laabs1996} and the $M^2$ factor increased. After 15\,s we reversed the process to return to the fundamental Gaussian mode after 30\,s. The reachable speed of the method is only limited by the used hardware.

\section{Conclusion}

We have shown that a modal decomposition of an investigated laser beam allows not only the quantitative determination of its transverse modal content but also of its beam quality factor $M^2$. Using the correlation filter technique, it is possible to determine the modal spectrum in real-time. We demonstrated the working capabilities by analyzing the beam emerging from an end-pumped Nd:YAG laser. By changing the alignment of its cavity, several Hermite-Gaussian modes of higher order could be excited. We showed that a correlation analysis involving the first 21 Hermite-Gaussian modes yields values for a derived $M^2$ parameter, which are in good agreement with the caustic measurements proposed by the ISO standard. The advantage of the method is the ability for real-time analysis of the beam that was demonstrated by continuously changing the alignment of the laser cavity and monitoring the beam propagation ratio $M^2$ as a function of time.

We conclude that this method provides an experimentally simple possibility to measure the beam propagation ratio $M^2$, which is compatible to the ISO standard in cases where caustic measurements are too time-consuming or even impossible to perform. The additional information about the modal spectrum of the beam is highly advantageous for diverse laser applications.


\begin{thebibliography}{99}

\bibitem{Siegman1998} 
A.~E. Siegman, ``How to (maybe) measure laser beam quality,'' in \emph{DPSS (Diode Pumped Solid State) Lasers: Applications and Issues}, M. Dowley, ed., Vol. 17 of OSA Trends in Optics and Photonics (Optical Society of America, 1998), paper MQ1.

\bibitem{ISO11146}
{International Organization for Standardization}, \emph{ISO 11146-1/2/3 {T}est
  methods for laser beam widths, divergence angles and beam propagation ratios
  -- Part 1: Stigmatic and simple astigmatic beams / Part 2: General
  astigmatic beams / Part 3: Intrinsic and geometrical laser beam
  classification, propagation and details of test methods} (ISO, Geneva, 2005).

\bibitem{Schafer2002}
B.~Sch\"{a}fer and K.~Mann, ``Determination of beam parameters and
  coherence properties of laser radiation by use of an extended
  Hartmann-Shack wave-front sensor,'' Appl. Opt. \textbf{41}, 2809--2817
  (2002).

\bibitem{Eppich2005}
B.~Eppich, G.~Mann, and H.~Weber, ``Measurement of the four-dimensional
  Wigner distribution of paraxial light sources,''  in \emph{Optical Design and Engineering II}, 
  L.~Mazuray and R.~Wartmann, eds., Proc. SPIE \textbf{5962}, 59622D (2005).

\bibitem{Lambert2004}
R.~W. Lambert, R.~Cort\'{e}s-Mart\'{i}nez, A.~J. Waddie, J.~D. Shephard, M.~R.
  Taghizadeh, A.~H. Greenaway, and D.~P. Hand, ``Compact optical system
  for pulse-to-pulse laser beam quality measurement and applications in laser
  machining,'' Appl. Opt. \textbf{43}, 5037--5046 (2004).

\bibitem{Siegman1986}
A.~E. Siegman, \emph{Lasers} (University Science Books, Sausalito, 1986).

\bibitem{Saghafi1998}
S.~Saghafi and C.~J.~R. Sheppard, ``The beam propagation factor for
  higher order Gaussian beams,'' Opt. Commun. \textbf{153}, 207--210 (1998).

\bibitem{Tervonen1989}
E.~Tervonen, J.~Turunen, and A.~Friberg, ``Transverse laser mode
  structure determination from spatial coherence measurements: Experimental
  results,'' Appl. Phys. B \textbf{49}, 409--414 (1989).

\bibitem{Cutolo1995}
A.~Cutolo, T.~Isernia, I.~Izzo, R.~Pierri, and L.~Zeni, ``Transverse
  mode analysis of a laser beam by near-and far-field intensity measurements,''
  Appl. Opt. \textbf{34}, 7974--7978 (1995).

\bibitem{Santarsiero1999}
M.~Santarsiero, F.~Gori, R.~Borghi, and G.~Guattari, ``Evaluation of the
  modal structure of light beams composed of incoherent mixtures of
  Hermite-Gaussian modes,'' Appl. Opt. \textbf{38}, 5272--5281 (1999).

\bibitem{Xue2000}
X.~Xue, H.~Wei, and A.~G. Kirk, ``Intensity-based modal decomposition of
  optical beams in terms of Hermite-Gaussian functions,'' J. Opt. Soc. Am. A
  \textbf{17}, 1086--1091 (2000).

\bibitem{Andermahr2008}
N.~Andermahr, T.~Theeg, and C.~Fallnich, ``Novel approach for
  polarization-sensitive measurements of transverse modes in few-mode optical
  fibers,'' Appl. Phys. B \textbf{91}, 353--357 (2008).

\bibitem{Soifer1994}
V.~A. Soifer and M.~Golub, \emph{Laser Beam Mode Selection by Computer
  Generated Holograms} (CRC Press, Boca Raton, 1994).

\bibitem{Duparre2005}
M.~Duparr\'{e}, B.~L\"{u}dge, and S.~Schr\"{o}ter, ``On-line
  characterization of Nd:YAG laser beams by means of modal decomposition
  using diffractive optical correlation filters,'' in \emph{Optical Design and Engineering II}, 
  L.~Mazuray and R.~Wartmann, eds., Proc. SPIE \textbf{5962}, 59622G (2005).

\bibitem{Kaiser2009a}
T.~Kaiser, D.~Flamm, S.~Schr\"{o}ter, and M.~Duparr\'{e}, ``Complete
  modal decomposition for optical fibers using CGH-based correlation
  filters,'' Opt. Express \textbf{17}, 9347--9356 (2009).

\bibitem{Flamm2010a}
D.~Flamm, O.~A. Schmidt, C.~Schulze, J.~Borchardt, T.~Kaiser, S.~Schr\"{o}ter,
  and M.~Duparr\'{e}, ``Measuring the spatial polarization distribution
  of multimode beams emerging from passive step-index large-mode-area fibers,''
  Opt. Lett. \textbf{35}, 3429--3431 (2010).

\bibitem{Kogelnik1966}
H.~Kogelnik and T.~Li, ``Laser beams and resonators,'' Appl. Opt.
  \textbf{5}, 1550--1567 (1966).

\bibitem{Laabs1996}
H.~Laabs and B.~Ozygus, ``Excitation of Hermite-Gaussian modes in
  end-pumped solid-state lasers via off-axis pumping,'' Optics \& Laser
  Technology \textbf{28}, 213--214 (1996).

\bibitem{Szego1975}
G.~Szeg\"o, \emph{Orthogonal Polynomials} (Amer. Math. Soc., Providence, 1975).

\end{thebibliography}
\end{document}